\documentclass[aps,pre,preprint,showpacs,groupedaddress,amssymb]{revtex4}

\usepackage{graphicx} 

\begin{document}
\title{Supersonic dislocations observed in a plasma crystal}
\author{V.~Nosenko, S.~Zhdanov, G.~Morfill}
\affiliation {Max-Planck-Institut f\"{u}r extraterrestrische Physik,
D-85741 Garching, Germany}
\date{\today}
\begin{abstract}
Experimental results on the dislocation dynamics in a
two-dimensional plasma crystal are presented. Edge dislocations were
created in pairs in lattice locations where the internal shear
stress exceeded a threshold and then moved apart in the glide plane
at a speed higher than the sound speed of shear waves, $C_T$. The
experimental system, a plasma crystal, allowed observation of this
process at an atomistic (kinetic) level. The early stage of this
process is identified as a stacking fault. At a later stage,
supersonically moving dislocations generated shear-wave Mach cones.
\end{abstract}
 \maketitle

Dislocations are ubiquitous in solids \cite{Frank:52}. They are
essential to understanding such properties as plasticity, yield
stress, susceptibility to fatigue, fracture, and
dislocation-mediated melting of 2D solids. Dislocation generation
and motion is therefore of interest in materials science
\cite{Kittel}, the study of earthquakes \cite{Archuletta:82}, snow
avalanches \cite{Kirchner:02}, colloidal crystals
\cite{Schall-Kolar}, 2D foams \cite{Kader:99}, and various types of
shear cracks \cite{Abraham:00,Rosakis:99}.

The theory of dislocations uses several approaches. In elastic
theory, a dislocation's core is treated as a singularity in an
otherwise continuous elastic material. Linear elastic theory
predicts that a gliding edge dislocation cannot overcome the sound
speed of shear waves $C_T$, because the energy radiated by a moving
dislocation becomes infinite at this speed. However, a gliding edge
dislocation can move at a particular speed of $\sqrt{2}C_T$
\cite{Eshelby:49}; in this case, it does not emit any radiation at
all and therefore its motion is frictionless. A more detailed theory
of dislocations should take into account the discreet ``atomistic''
structure of matter. In Ref.~\cite{Eshelby:49}, this was taken into
account in the dislocation's glide plane, outside, the material was
considered to be an elastic continuum. Then, the limiting speed for
dislocations is the Rayleigh wave speed $C_R$; usually, $C_R\lesssim
C_T$.

Gliding edge dislocations moving at the speed of $1.3C_T$ to
$1.6C_T$ were observed in atomistic computer simulations
\cite{Gumbsch:99}. Dislocations were created with a speed exceeding
$C_T$ at a strong stress concentration. High shear stress was
required to sustain their motion. Dislocations always radiated
strongly, even when they traveled at exactly $\sqrt{2}C_T$. This was
attributed to nonlinear effects in the dislocation core. Shear
cracks propagating faster than $C_T$ were observed in
Ref.~\cite{Rosakis:99}. However, to the best of our knowledge,
experimental evidence that dislocations can move faster than $C_T$
is lacking.

In regular solids dislocation dynamics is almost impossible to study
experimentally at an atomistic level \cite{Murayama:02} because of
the small distances between the atoms (or molecules), high
characteristic frequencies, and the lack of experimental techniques
of visualizing the motion of individual atoms or molecules.

The most suitable model systems to experimentally study dislocations
at an atomistic level are plasma crystals \cite{Lin_I:96}. These are
suspensions of highly charged micron-sized particles in a plasma,
which can be strongly coupled \cite{Thomas:94_etc}. Then their
mutual interaction causes them to self-organize in structures that
can have crystalline or liquid order. The interparticle distance can
be of the order of $100~\mu$m to $1$~mm, characteristic frequencies
are of the order of $10-100~\rm{s}^{-1}$, and the speed of sound is
of the order of $10$~mm/s. In addition, the absence of a substantial
background medium (as in the case of colloids) allows studies of the
full dynamics without overdamping. These unique characteristics,
plus the direct imaging, make it possible to study the complex
dynamics of crystalline defects \cite{Lin_I:04}, including
dislocation nucleation and motion \cite{Reichhardt:03,Lin_I:96},
all at an atomistic level.

Our experimental setup was a modified GEC chamber as in Ref.~\cite
{Nosenko:06LH}, using similar experimental parameters. Argon plasma
was produced using a capacitively-coupled rf discharge. We used
$42$~W of rf power at $13.56$~MHz, with an amplitude of $158$~V
peak-to-peak. The self-bias voltage was $-96$~V. To ensure that the
system was not overdamped, a relatively low pressure of $3$~mTorr
was used. The neutral-gas damping rate is then accurately modeled
\cite {Liu:03} by the Epstein expression $\nu=\delta
N_gm_g\overline{v}_g(\rho_pr_p)^{-1}$, where $N_g$, $m_g$, and
$\overline{v}_g$ are the number density, mass, and mean thermal
speed of gas atoms and $\rho_p$, $r_p$ are the mass density and
radius of the particles, respectively. With leading coefficient
$\delta=1.26$ \cite {Liu:03}, this gave $\nu=0.52$~s$^{-1}$.

A monolayer of highly charged microspheres was levitated against
gravity in the sheath above the lower rf electrode. The particles
had a diameter of $8.09\pm0.18$~$\mu$m \cite {Liu:03} and a mass
$m=4.2 \times 10^{-13}$~kg. The monolayer included $\approx 4500$
particles, had a diameter of $\approx60$~mm, and rotated slowly in
the horizontal plane.

\begin{figure*}
\centering
\includegraphics[width=136mm]{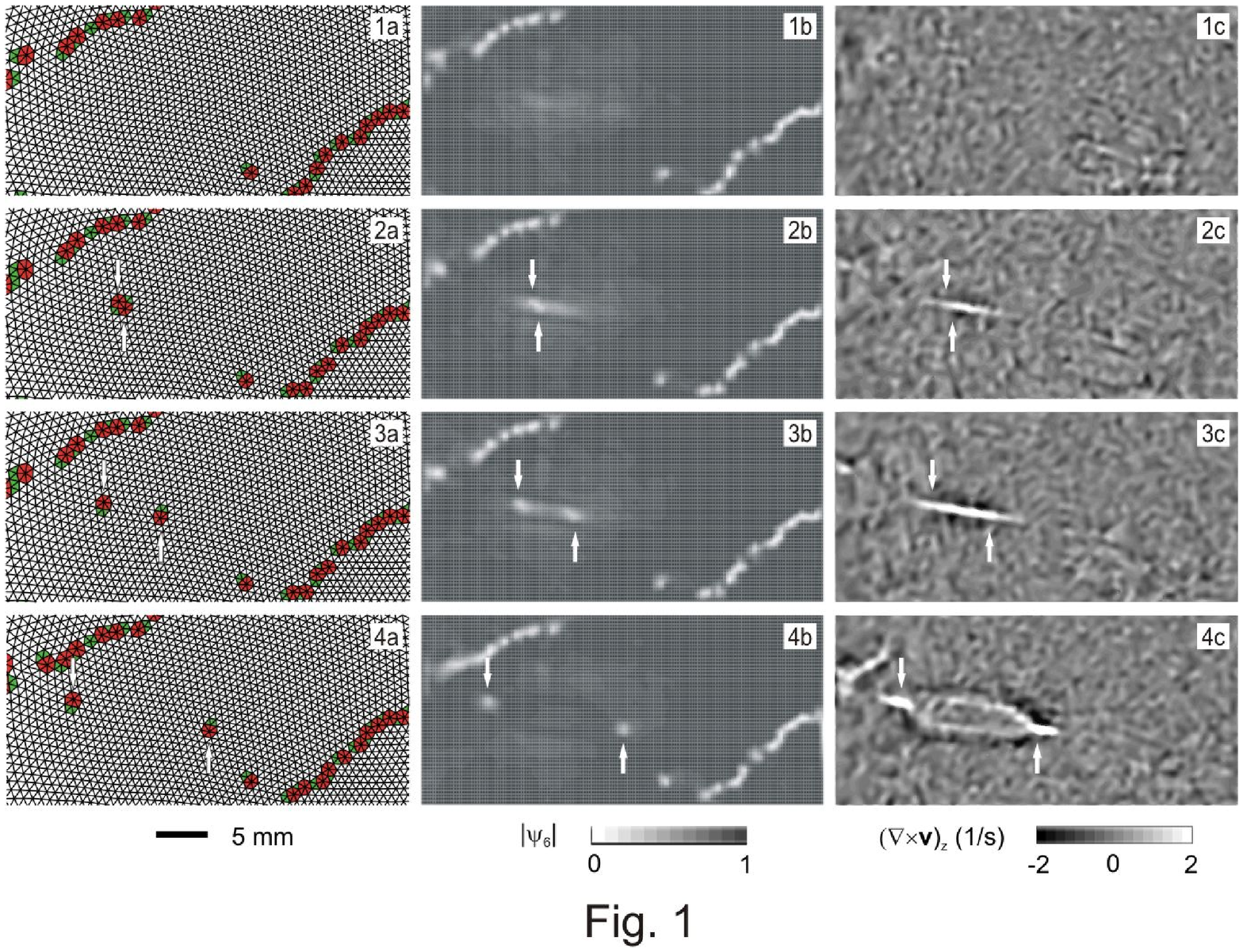}
\caption {\label {Maps_3x4} (Color online) Dislocation nucleation
and motion in a 2D crystalline lattice. Maps of (a) triangulation of
the particle positions, (b) bond-orientational function $|\psi_6|$,
and (c) vorticity $(\nabla \times \bf{v})_z$, where $\bf{v}$ is the
particle velocity, are shown for four different instants of time:
(1) $0.335$~s, (2) $0.569$~s, (3) $0.669$~s, and (4) $1.004$~s. A
pair of dislocations, indicated by arrows, was created shortly
before (2).}
\end{figure*}

The interparticle potential for particles arranged in a single
plane, like ours, is well approximated \cite {Konopka:00:Yukawa} by
the Yukawa potential: $U(r)=Q(4\pi\epsilon_0r)^{-1}{\rm
exp}(-r/\lambda_D)$, where $Q$ is the particle charge and
$\lambda_D$ is the screening length. The monolayer is characterized
by a screening parameter $\kappa=b/\lambda_D$, where $b$ is the
interparticle spacing. In our experiment, $b=0.69$~mm. We used the
pulse technique of Ref.~\cite {Nosenko:02}, making use of a
theoretical wave dispersion relation \cite {Wang:01,Zhdanov:03}, to
measure $\kappa=1.58\pm0.22$ and $Q=-15~000\pm1400e$. The average
sound speeds in the central part of our plasma crystal were measured
to be $C_L=24.0\pm1.7$~mm/s and $C_T=6.3\pm0.6$~mm/s, for
compressional and shear waves respectively. (Based on these
measurements, the Rayleigh wave speed is $C_R\approx6.0~{\rm
mm/s}=0.95C_T$, formally using the approach of
Ref.~\cite{Eshelby:49}).

The particles were imaged through the top window by a digital
camera. We recorded movies of $1024$ frames at $29.88$ frames per
second. The $61.1\times45.8~\rm{mm}^2$ field of view included
$\approx 4200$ particles. The particle coordinates $x,y$ and
velocities $v_x,v_y$ were then calculated with subpixel resolution
for each particle in each frame.

At our experimental conditions, the particle suspension
self-organized in an ordered triangular lattice with hexagonal
symmetry. This lattice always contained defects, as revealed by
Delaunay triangulation, Fig.~\ref {Maps_3x4}(a). A defect is defined
as a lattice site where a particle has a number of nearest neighbors
other than six. Defects are highlighted in Fig.~\ref {Maps_3x4}(a).
Most of them form linear chains that constitute domain walls in our
2D crystal. Two nearly parallel domain walls are seen in Fig.~\ref
{Maps_3x4}(a).

An edge dislocation in our 2D crystal consists of an isolated pair
of 5- and 7-fold defects. During the course of experiment,
dislocations in the lattice are continuously generated due to the
shear introduced by the slow rotation. They move around and
annihilate with each other or with domain walls. Some of them may
remain stationary for some time. In this Letter, we show that edge
dislocations are created in pairs in lattice locations where the
internal shear stress exceeds a threshold and then move apart
supersonically.

To study dislocation nucleation, we evaluate the shear strain in the
lattice from the bond-orientational function $|\psi_6|$ \cite
{Grier_Knapek}, shown in Fig.~\ref {Maps_3x4}(b). For every site in
the lattice, $\psi_{6}=\frac{1}{n}\sum^n_{j=1}{\rm
exp}(6i\Theta_{j})$, where $\Theta_{j}$ are bond orientation angles
for $n$ nearest neighbors. In the limit of weak \textit{simple
shear}, the following relation can be used: $|\psi_6|=1-9\gamma^2$,
where $\gamma$ is the shear strain. For weak \textit{pure shear},
$|\psi_6|=1-2.25e^2$, where $e$ is elongation, which is the measure
of pure shear deformation. We derived these relations assuming small
deformations of an elementary hexagonal cell. Note that $|\psi_6|$
is insensitive to uniform compressions, rotations, and translations,
i.e., the deformations that do not involve any variation of bond
orientations inside an elementary cell.

To study dislocation dynamics, we use the maps of vorticity $(\nabla
\times \bf{v})_z$, Fig.~\ref {Maps_3x4}(c), where $\bf{v}$ is the
particle velocity ($z$ denotes the out-of-plane component of
vorticity). These reveal shear motion ~\cite {Nosenko:02,Nosenko:03}
and are therefore suitable for visualizing moving dislocations.

The shear strain in the lattice had a non-uniform distribution, as
follows from Fig.~\ref {Maps_3x4}(1b). The shear strain was higher
(or $|\psi_6|$ lower) in two kinds of locations. First, it was high
in domain walls - the two nearly-parallel bright stripes in
Fig.~\ref {Maps_3x4}(b) (or equivalently the chains of 5- and 7-fold
defects in Fig.~\ref {Maps_3x4}(a)). Second, a ``diffuse''
concentration of shear strain appeared between the domain walls. We
attribute this diffuse strain concentration to the differential
rotation and shear in the lattice, with two ``rigid'' domain walls
imbedded in it.

The diffuse shear strain increased with time. When it locally
exceeded a certain threshold, a pair of edge dislocations was
created in that location, Fig.~\ref {Maps_3x4}(2). These
dislocations appear as bright spots in Fig.~\ref {Maps_3x4}(2b) or
as pairs of 5- and 7-fold defects in Fig.~\ref {Maps_3x4}(2a), all
indicated by arrows. Once a pair of dislocations was created, they
moved rapidly apart, Figs.~\ref {Maps_3x4}(3),\ref {Maps_3x4}(4).

\begin{figure}
\centering
\includegraphics[width=70mm]{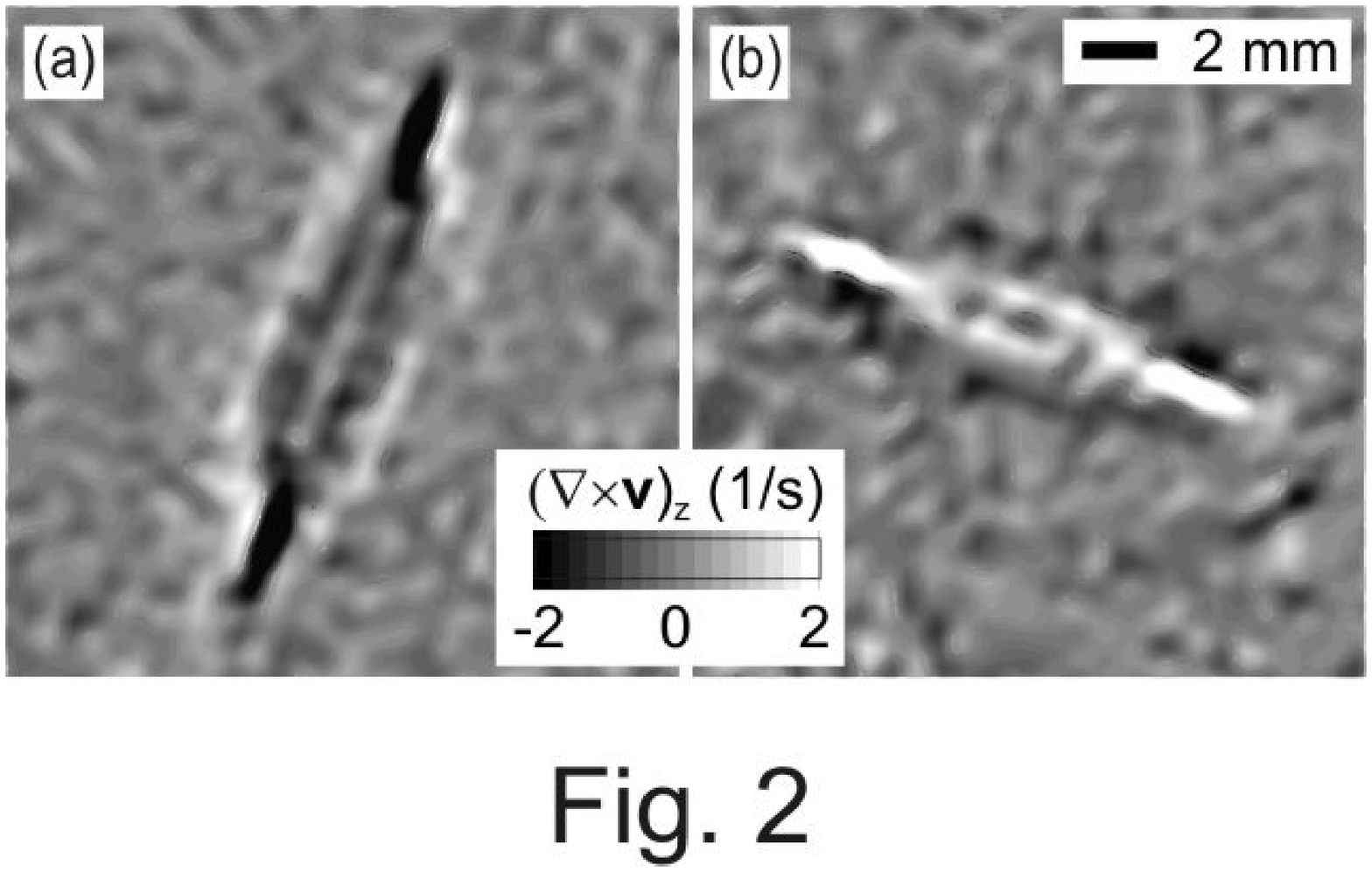}
\caption {\label {More_Examples} More examples of dislocation
nucleation similar to that in Fig.~\ref {Maps_3x4}(4c). In (a)
``black'' and (b) ``white'' dislocation cores, particles rearrange
in the clockwise and counterclockwise directions, respectively.}
\end{figure}

During the course of our $5.7$~min experiment we observed about $30$
events of dislocation nucleation similar to that shown in Fig.~\ref
{Maps_3x4}. Two more typical examples are shown in Fig.~\ref
{More_Examples}. The ``black'' and ``white'' dislocations have
opposite directions of the particle rearrangement in their cores.

\begin{figure}
\centering
\includegraphics[width=70mm]{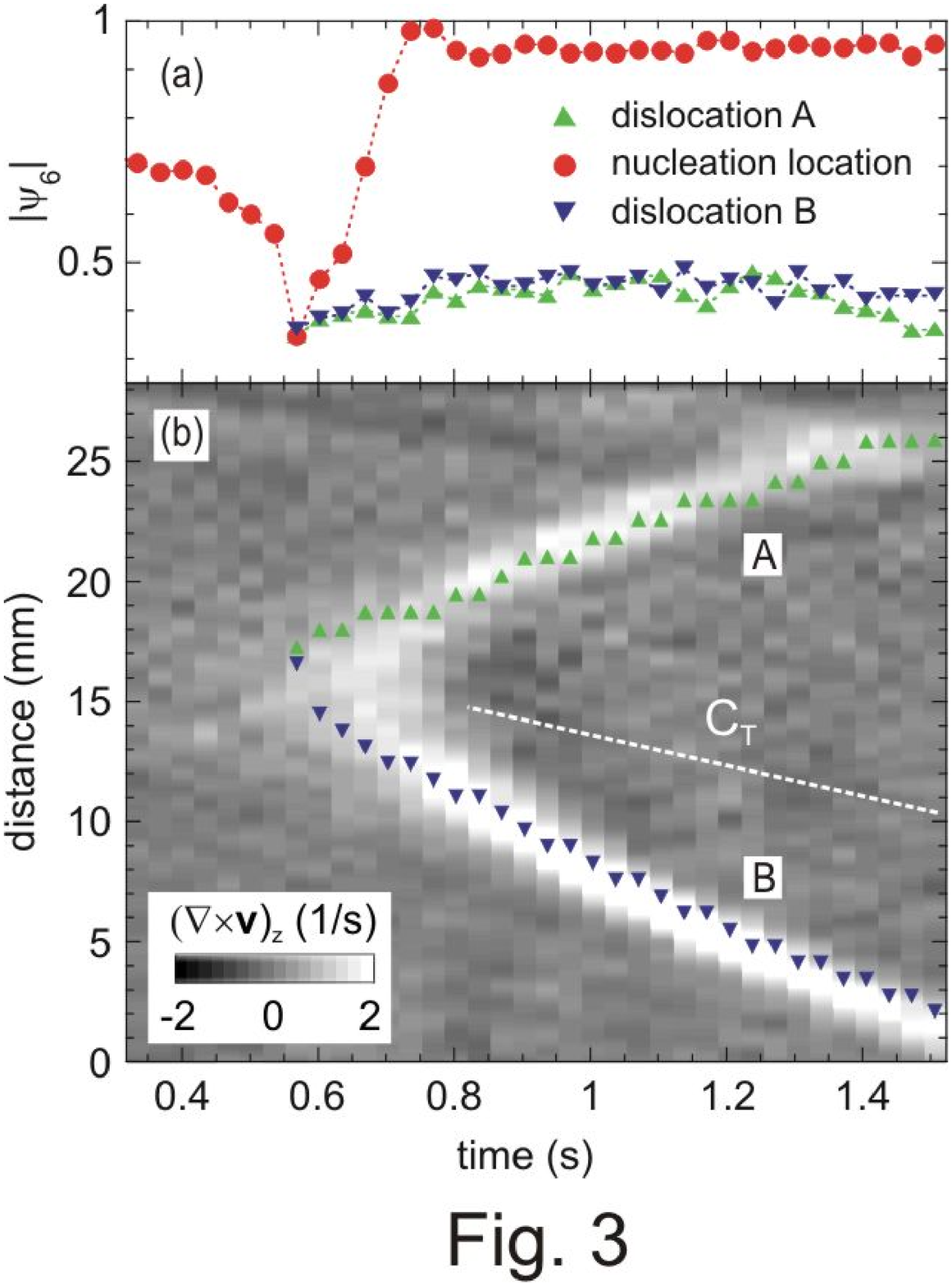}
\caption {\label {Strain_n_V} (Color online) Pair of dislocations is
generated and moves apart. (a) Time evolution of the
bond-orientational function $|\psi_6|$. We use $|\psi_6|$ to
evaluate the shear strain, which is higher when $|\psi_6|$ is lower;
for details, see the text. (b) The grey-scale space-time plot shows
vorticity $(\nabla \times \bf{v})_z$ measured along the
dislocations' glide plane. Superimposed are the positions of
dislocation cores calculated as average positions of the respective
5- and 7-fold lattice defects. Dislocations $A$ and $B$ are
respectively LHS and RHS in Fig.~\ref {Maps_3x4}.}
\end{figure}

We examine the dislocation nucleation and dynamics in more detail.
The time evolution of the shear strain in the lattice location where
a pair of dislocations is generated has several stages, as shown in
Fig.~\ref {Strain_n_V}(a). First, the shear strain builds up
gradually in a certain location ($|\psi_6|$ decreases). Second, when
the shear strain in this location exceeds a certain threshold, a
pair of dislocations is born. Third, the shear stress is rapidly
relieved when the dislocations separate, and gradually drops to the
background value. This cycle then starts over again, perhaps in a
different location.

When a pair of dislocations is created, their Burgers vectors are
oppositely directed and equal in magnitude, within the experimental
errors and inhomogeneity of the lattice. Therefore, the total
Burgers vector is conserved in this process. (The Burgers vector
represents the magnitude and direction of the crystalline lattice
distortion by a dislocation.)

When the two dislocations separate, they leave a stacking fault
between them. It appears as a narrow (one lattice constant) band
where the lattice structure is distorted from triangular to nearly
square, Fig.~\ref {Maps_3x4}(3a). This stacking fault has a dynamic
nature; the lattice is restored to its original state by rapid
particle rearrangement that is seen as shear motion in Fig.~\ref
{Maps_3x4}(3c). A similar shear motion occurs even ahead of the
right-hand-side dislocation in Fig.~\ref {Maps_3x4}(2c).

Next, we analyze the speed of the dislocations as they move apart.
Fig.~\ref {Strain_n_V}(b) shows a grey-scale space-time plot of
vorticity $(\nabla \times \bf{v})_z$ measured along the
dislocations' glide plane. Superimposed are the positions of
dislocation cores. Note that the dislocation motion is not smooth;
rather, it is of a stick-and-slip type. The average dislocation
speeds are higher than $C_T$; however, they decrease as dislocations
move. (The slope corresponding to $C_T$ is shown by a dashed line in
Fig.~\ref {Strain_n_V}(b)). For comparison, the linear shear
disturbance in the upper part of Fig.~\ref {Strain_n_V}(b) (seen as
a faint black feature) travels at a speed $\approx C_T$. The average
speeds of dislocations $A$ and $B$ are respectively $U_A=9.5$~mm/s
and $U_B=13$~mm/s. We attribute this difference to the radial
variation of the local number density of our crystal. It diminished
from $2.4~{\rm mm}^{-2}$ in the crystal center ($B$ moved in this
direction) to $\approx 1.6~{\rm mm}^{-2}$ at its periphery where we
still observed dislocations ($A$ moved here).

\begin{figure}
\centering
\includegraphics[width=60mm]{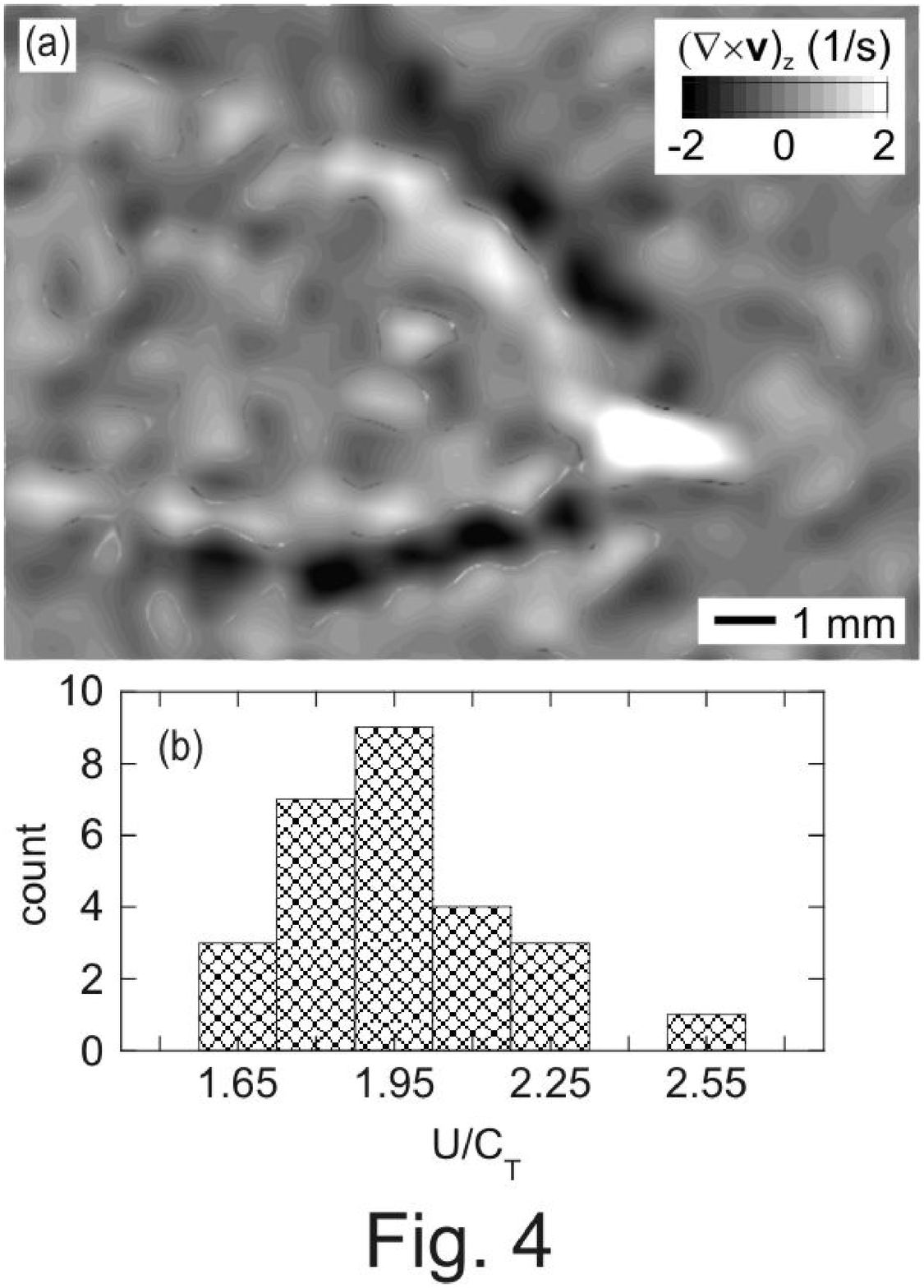}
\caption {\label {Mach_cone} (a) Shear-wave Mach cone generated by a
supersonically moving dislocation $B$ from Fig.~\ref {Strain_n_V},
at $t=1.406$~s. (b) Distribution of supersonic dislocation speeds
$U$.}
\end{figure}

Dislocations that move supersonically create a distinct signature,
i.e., a shear-wave Mach cone, shown in Fig.~\ref {Mach_cone}(a). The
structure is similar to that observed in
Refs.~\cite{Gumbsch:99,Nosenko:02,Nosenko:03}. A Mach cone is a
V-shaped wake created by a moving supersonic disturbance. Mach cones
obey the Mach cone angle relation ${\rm sin}\,\mu=C/U$, where $\mu$
is the cone's opening angle, $C$ is the speed of sound, and $U$ is
the speed of the supersonic disturbance. From their Mach cone
angles, we derive the relative speeds of dislocations $A$ and $B$ in
Fig.~\ref {Strain_n_V}, $U_A=1.7C_T$ and $U_B=1.9C_T$. Since we
measured $U_A=9.5$~mm/s and $U_B=13$~mm/s by defect tracking, we can
estimate the \textit{local} sound speeds at the locations of $A$ and
$B$ as $C_{T,A}=5.5$~mm/s and $C_{T,B}=6.7$~mm/s, respectively.
Hence, $C_T$ was higher in the crystal center and somewhat declined
toward its periphery due to the corresponding decline in the
crystal's number density. A weak dependence of $C_T$ on number
density (or interparticle spacing) was discussed in
Ref.~\cite{Wang:01}. The Mach cones created by a dislocation pair
were connected by two parallel fronts of shear waves, Fig.~\ref
{Maps_3x4}(4c). This wave front configuration evolved from the
particle rearrangement in the stacking fault, Fig.~\ref
{Maps_3x4}(3c). The Mach cones were composed of shear waves and not
of compressional waves, because they were excited by dislocations
moving faster than $C_T$, but slower than $C_L$. Note that we did
not use any external exciter as in
Refs.~\cite{Melzer:00,Nosenko:02,Nosenko:03}; also there were no
particles above or beneath the monolayer as in
Ref.~\cite{Samsonov:99MC}.

Finally, we calculated the distribution of dislocation speeds,
Fig.~\ref {Mach_cone}(b), using the Mach cone angle relation and
measuring the opening angles of well-developed Mach cones in maps
like that in Fig.~\ref {Mach_cone}(a). From this distribution, we
conclude that the average speed of supersonic dislocations in our
experiment was $(1.95\pm0.2)C_T$, where $C_T$ is the sound speed of
shear waves.

We thank J.~Goree and T.~Sheridan for their help in conducting the
experiment.


\end{document}